\documentclass[aps,preprint]{revtex4}
\usepackage{epsfig}
\usepackage{amsmath}
\usepackage{epsfig}

\begin{document}
\title
{\bf Special rectangular (double-well and hole) potentials}    
\author{Zafar Ahmed$^1$, Tanayveer Bhatia$^2$, Shashin Pavaskar$^3$, Achint Kumar$^4$}
\email{1: zahmed@barc.gov.in, 2: tanayveer1@gmail.com,  3: spshashin3@gmail.com, 4: achint1994@gmail.com}
\affiliation{$^1$Nuclear Physics Division, Bhabha Atomic Research Centre,Mumbai 400 085, India.\\ $^2$Birla Institute of Technology and Science, Pilani, 333031, India.\\ $^3$National Institute of Technology, Surathkal, Mangalore, 575025, India.\\ $^4$Birla Institute of Technology and Science, Pilani, Goa, 403726, India.}
\date{\today}
\begin{abstract} 
We revisit a rectangular barrier as well as a rectangular well (pit) between two rigid walls. The former is the well known double-well potential and the latter is a hole potential. Let $|V_0|$ be the height (depth) of the barrier (well) then for a fixed geometry of the potential, we show that in the double-well, $E=V_0(>0)$,  and in the hole potential ($V_0 <0$), $E=0$, can be energy  eigenvalues provided $V_0$ admits some special discrete values. These  states have been missed out earlier which emerge only when one seeks the special zero-energy solution of one-dimensional Schr{\"o}dinger equation as $\psi(x)=Bx+C$. 
\end{abstract}
\maketitle
One dimensional Schr{\"o}dinger equation for free particle (zero potential) 
\begin{equation}
{d^2 \psi(x) \over dx^2}+k^2 \psi(x)=0, \quad  k=\sqrt{2mE \over \hbar^2},
\end{equation}
admits
\begin{equation}
\psi(x)= B \sin kx+ C \cos kx,
\end{equation}
and separately [1,2]  for $k=0$
\begin{equation}
\psi(x)=B x+C
\end{equation} 
as solutions.
Note that for $k=0$, (2) becomes trivially a constant. 
The solution (3) may also not satisfy the boundary condition  and fail to become an eigenstate. Like in the case of infinitely deep potential [1,2], it (3) vanishes identically  while satisfying the boundary condition at $x=\pm a$: $\psi(\pm a)=0$. However, a
surprising existence of such a zero-energy and zero-curvature eigenstate in the presence of Dirac delta potential has been 
revealed rather late [3]. Later investigations [4,5] show that the specially designed potentials  can give rise to eigenstates  having zero-curvature in  part(s) of the domain of the potential. 

Recently, it has been shown [6] that when the Dirac delta
potential $V(x)=V_0 \delta (x)$ is placed symmetrically between two rigid walls, the zero-energy
zero-curvature eigenstate does not exist when $V_0>0$ (double well potential) and it exists critically as a genuine ground state energy when Dirac delta potential lies symmetrically beneath the infinitely deep well ($V_0 < 0$) and ${mV_0 a \over \hbar^2}=-1$. Such a state has been missed out in discussions on this interesting eigenvalue problem (Problem nos. 19 and 20 in Ref. [7]).
  
The question arising here is whether the existence  of such a state has been investigated in simple potential models discussed in textbooks [7-10]. Here,
we show that  $E=V_0$ in the simple double well potential (rectangular barrier between two rigid walls, see the black line in Fig. 1) and $E=0$ in the potential hole (the rectangular well between two rigid walls, see the gray line in Fig. 1) can be the discrete 
energy eigenvalues provided the potential parameters satisfy  critical conditions. These special states  remain generally elusive in the discussions about these potentials in the textbooks [7-10]. These states are the consequence of the special zero-energy solution (3) of one-dimensional Sch{\"o}dinger equation which has been generally spared in these eigenvalue problems.

First let us place the Dirac delta potential $V(x)=V_0 \delta(x)$ non-symmetrically between two rigid walls at $x=-a$ and $x=c$. See the faint vertical line depicting Dirac Delta Potential at $x=0$ in Fig. 1. We look for the possibility of a zero-energy eigenstate for $V_0>0$. We can write the appropriate solution of (1) with regard to  (3) as: $\psi_<(x)=A(x+a),~ \mbox{for}~ -a<x<0$ and $\psi_>(x)=B(x-c) ~\mbox{for}~ 0<x<c.$ Let us match these two wave functions at $x=0$ to get $Aa=B(-c)$. Due to
the presence of Dirac delta the derivative of these wave functions will mismatch [7,11,12] at $x=0$ to give $B-A={2mV_0 \over \hbar^2}Aa$. Elimination of $A$ and $B$ from the last two equations gives 
\begin{equation}
{2mV_0 a \over \hbar^2}=-\left (1+{a \over c}\right). 
\end{equation}
We conclude that  only negative values of $V_0$ can allow $E=0$ to become an eigenstate provided the condition (4) is met. This is why the double well potential made of delta barrier placed symmetrically [6] or non-symmetrically can not admit zero-energy and zero-curvature eigenstate.

The simple double well potential is an often-discussed problem in textbooks [8-10]. However, while discussing its eigenvalue problem, $E = 0,V_0$, have either not been checked to be an eigenvalue or they have been discarded, as the case-specific inappropriate solution (2) has been forced in the region $x \in [-b,b]$. Here we use the appropriate solution (3) to detect that $E=0,V_0$ can become eigenvalues of the potential  as depicted in Fig. 1. The double well or the hole  potential (Fig. 1) is written as
\begin{equation} 
V(x)=\left\{ \begin{array}{lcr}
\infty, & & x\le -a, x\ge c\\
0, & & -a <x<-b, b<x<c\\
V_0, & & -b\le x\le b
\end{array}
\right.
\end{equation}
We propose to solve the discrete eigenvalue problem
for this potential in (1). We consider three separate cases (i): When $E=0$, (ii) $E\ne 0, V_0$, (iii) $E=V_0$.
In the literature only case (ii) is discussed
mostly for a symmetric double well [8-10]. The symmetric case of the hole potential  ($V_0<0$) has
also been discussed (see Problem no. 26 in Ref. [7])
\\ \\
{\bf Case (i): Boundstate at}$ {\bf ~ E=0}$\\ \\
The Schr{\"o}dinger equation in zero-potential region is (1)
and for zero-energy we seek its solution as
\begin{equation}
\begin{array}{lcr}
\psi_<(x)=A(x+a),& &  x\in [-a,-b]\\ \psi_>(x)=D(x-c),& &  x\in [b,c]
\end{array}
\end{equation}
which are compatible and vanish at their respective boundaries at $x=-a$ and $x=c$.
In the barrier region for zero-energy the Schr{\"o}dinger equation is
\begin{equation}
{d^2 \psi(x) \over dx^2}-q^2 \psi(x)=0, \quad   \forall~ x  \in [-b,b] \quad \mbox{, where} \quad q=\sqrt{2mV_0 \over \hbar^2}
\end{equation}
whose solution is
\begin{equation}
\psi(x)=B \sinh qx+ C \cosh qx.
\end{equation}
Matching the wave functions and their derivatives at $x=-b$
we get
\begin{equation}
\begin{array}{c}
A(a-b)=-B \sinh qb+C \cosh qb\\
A=qB \cosh qb-qC \sinh qb 
\end{array}
\end{equation}
Similarly at $x=b$ we get
\begin{equation}
\begin{array}{c}
B \sinh qb + C \cosh qb= D(b-c)\\
qB \cosh qb + qC \sinh qb =D
\end{array}
\end{equation}
Let us  introduce  $a-b=d_1, c-b=d_2.$
In order to get the quantization condition or eigenvalue formula for $E$, one has to eliminate
$A,B,C$ and $D$ from these equations. One can find the ratio $B/C$ from Eqs. (8,9) and equate them to get
the energy eigenvalue equation. This requires a careful handling of denominators involving discontinuous functions $\tan \theta$ and $\cot \theta$  in various cases. We use the  simplest and the most general method (see Ref. [13] for the square well potential) to treat Eqs. (8,9) as linear simultaneous homogeneous equations of $A,B,C,D$ and look for their non-trivial ($A,B,C,D \ne 0$) solutions (see the  Appendix). This method in our present case, demands that:
\begin{equation}
\left |\begin{array} {cccc} -\sinh qb & 
\cosh qb & 0 & d1 \\  q\cosh qb & -q \sinh qb & 0 & 1\\  \sinh q b & \cosh q b  & d_2 & 0 \\ q \cosh qb & q \sinh qb & -1 & 0 \\
\end{array} \right |=0.
\end{equation}
Upon simplification we find the condition on the potential parameter for the existence of zero-energy eigenstate in the double well potential as:
\begin{equation}
(d_1 + d_2) q \cosh 2qb + (1 + d_1 d_2 q^2) \sinh 2qb=0
\end{equation}

Eventually, as each and every term in the above expression is positive definite, this equation cannot have real roots of $q$ for fixed values of the widths $d_1$ and $d_2$.

When we change $V_0$ to $-U$, q $\rightarrow$  $ i r$  and the hyperbolic functions become trigonometric and Eq. (12) becomes
\begin{equation}
(d_1 + d_2) r \cos 2br + (1 - d_1 d_2 r^2) \sin 2br=0,\quad where \quad r=\sqrt{2mU \over \hbar^2}.
\end{equation}
This trigonometric implicit equation will have infinitely many roots for $r(U_n)$, where the first root will give the depth $(U_0)$ of the well so as to admit $E=0$ as ground state. Then next roots $U_n$ will 
give values of depth so as to admit $E=0$ as some $n^{th}$ excited state of the total potential.\\

{\bf Case (ii): Boundstate at} ${\bf E \ne 0, V_0}$\\ \\
Now we work out the usual [8-10]  non-zero eigenvalues of the potential given in (5). Instead
of the solutions (6) we will now have for $E \ne 0$.
\begin{equation}
\begin{array}{lcr}
\psi_<(x)=A \sin k(x+a),& & x \in [-a,-b]\\
\psi_>(x)=D \sin k(x-c),& & x \in [b,c]
\end{array}
\end{equation}
For the region $-b<x<b$ we have
\begin{equation}
\psi(x)=B \sinh p x + C \cosh p x,\quad \mbox{where} \quad p=\sqrt{2m(V_0-E) \over \hbar^2}. 
\end{equation}
We match the solutions and their derivatives at $x= -b$, we get
\begin{equation}
\begin{array}{c}
A \sin k(a-b) = -B \sinh p b + C \cosh p b\\
k A \cos k(a-b)= p B \cosh p b-p C \sinh p b
\end{array}
\end{equation}
Similarly, the matching conditions at $x=b$ give
\begin{equation}
\begin{array}{c}
 B \sinh p b + C \cosh p b= D \sin k(b-c)\\
p B \cosh p b+p C \sinh p b =D k\cos k(b-c)
\end{array}
\end{equation}
Again we demand the consistency of the above four Eqs. (16,17) and their non-trivial solutions for $A,B,C,D$, we get 
\begin{equation}
\left |\begin{array} {cccc} -\sinh pb & 
\cosh pb & 0 & \sin kd_1  \\ p \cosh pb & -p \sinh p b & 0 & k \cos kd_1 \\ \sinh pb & \cosh p b & \sin kd_2 & 0 \\  p \cosh pb & p \sinh pb & -k \cos k d_2 & 0
\end{array} \right |=0.
\end{equation} 
The condition (18) simplifies to:
\begin{equation}
\begin{array}{c}
\pm\sqrt{E(V_0-E)}~ \cosh 2pb ~\sin k d + [E \cos kd + V_0 \sin kd_1~ \sin kd_2]~ \sinh 2bp=0\\
\mbox {for}  \quad E \ne 0 \quad \mbox{where} \quad d=d_1+d_2.
\end{array}
\end{equation}
In above Eq. (19), the + (-) sign is to be taken for $E>0(<0)$
It needs to be remarked that the Eqs. (6-13) can not be obtained as a limiting case of Eqs. (14,19). Although the Eqs. (18,19) are for $E\ne 0$, yet $k = 0$ satisfies them  spuriously. It can however be readily checked that both Eqs. (16) and (17) separately yield $B=0=C$.
Thus, the already discussed zero-energy condition (12) on the potential parameters and the eigenfunctions
in (6,8) would lead to the correct treatment of $E=0$ for the hole potential. Similarly, $E=V_0$ satisfies Eq. (19) un-intently
and this energy requires a separate treatment which is given below. 
\\ 

{\bf Case (iii): Boundstate at} ${\bf E =V_0>0}$\\ \\
Here, we define $s=\sqrt{2mV_0 \over \hbar^2}$ and the solution of (1)
\begin{equation}
\begin{array}{lcr}
\psi_<(x)=A \sin s(x+a),& & x\in[-a,-b]\\
\psi_>(x)=D \sin s(x-c),& & x\in[b,c]
\end{array}
\end{equation}
For the region $-b<x<b$ we have
\begin{equation}
\psi(x)=B x + C.
\end{equation}
We match the solutions and their derivatives at $x= -b$, we get
\begin{equation}
\begin{array}{c}
A \sin s(a-b) = -B b + C \\
s A \cos s(a-b)= B.
\end{array}
\end{equation}
Similarly, the matching conditions at $x=b$ give
\begin{equation}
\begin{array}{c}
B b + C = D \sin s(b-c)\\B = D s\cos s(b-c)
\end{array}
\end{equation}
Once again the consistency condition for nontrivial
solutions of $A,B,C,D$ arising from Eqs. (22,23)
yields
\begin{equation}
\left |\begin{array} {cccc}  -b & 
1 & 0 & \sin sd_1  \\ 1 & 0 & 0 & s \cos sd_1 \\  b & 1 & \sin sd_2 & 0 \\ 1 & 0 & -s \cos s d_2 & 0 \\
\end{array} \right |=0.
\end{equation} 
Upon expansion of this determinant we get\\
\begin{equation}
2bs^2 \cos sd_1 \cos s d_2+ s \sin s(d_1+d_2)=0.
\end{equation}
We would like to emphasize here again that Eqs. (20-25) 
can not degenerate as a limiting case from Eqs. (14-19). We find the roots of this equation to get 
the special values of $V_0$ for various values of the parameters: $V_0, a,b,c$ or $V_0,b, d_1, d_2$.
For the symmetric case ($d_1=d=d_2$), we get that $s$
is a root of the simpler equation given as $\tan sd + bs=0$. 

For calculations, we take $2m=1=\hbar^2$, fix $a=3, b= 1, c=3$ ($d_1=2, d_2=2$) and vary $V_0$ to determine the first six eigenvalues of the potential using Eq. (19). First, we take $V_0=0$ and calculate eigenvalues which turn up as the well known eigenvalues, $E_n={n^2 \pi^2 \over 36}$,  of the infinitely deep well of width 6 unit. Then we take $V_0=10$ to appreciate the well known [8-10] characteristic sub-barrier doublets of eigenvalues.
See the Table I for the first six discrete energy eigenvalues. We also study the cases of $V_0=0,10$ for asymmetric case when $a=2,b=1,c=3$ (see Table II). For $V_0=0$ the discrete eigenvalues of infinitely deep well of width 5 units are recovered as $E_n={n^2 \pi^2 \over 25}$. Interestingly, in the asymmetric double-well the closely lying sub-barrier doublets of energy eigenvalues have  disappeared. 
This may not be a common experience. We, however, find that asymmetry in a double well may cause increased gap in successive even-odd pairs of eigenvalues compared to the case of a symmetric double well as  displayed here.

The cases of $V_0=-5.0$ (hole potential) in both Tables I and II display an ordinary spectrum wherein 
the first two levels are in the the hole with the next four positive energy levels in the box of width 
($c+a=d_1+d_2+2b$).    

Now we study the special $E=0$ eigenvalues using
the allowed discrete values of $V_0$ from Eqs. (12). 
For the symmetric case ($d_1= d= d_2= 2 ,b= 1$), we get $V_0=-0.4267, -3.3730, -10.8393, -23.1923$ as first four roots the trigonometric Eq. (12). $E_0=0$ is the ground state eigenvalue of (5)
when $0.4267$ is the depth of the hole, then $E_1=0$ is the first excited eigenvalue when the depth of the hole is $3.3730$, so on and so forth $E_2=0=E_3=E_4=E_5=0$ are the excited states of four potentials plotted in Fig. 2. Notice that zero-energy eigenstates (say,$\psi_Z(x)$) have the linear part $(Bx+C)$ for $|x|\le 1$. The other 5 discrete energy eigenvalues for these four potentials are available in Table I. We have checked that each $\psi_Z(x)$ is indeed orthogonal
\begin{equation}
\int_{-a}^{c} \psi_Z(x) \psi_n(x) dx =0.
\end{equation}
to all the 5 other  (listed in Table I) eigenstates of the same potential.  
 
In order to check the robustness of the zero-energy energy eigenstates we change $V_0$ to a value -0.5 (slightly different from the special value -0.4267) to increase the depth of the potential hole. Notice in the Table I that all the levels have been pushed down slightly with ground state at $E_0=-0.0497$ down
but close to zero. Similarly the change of $V_0$ from the special 
value $-3.3730$ to $-3$ to reduce the depth of the hole. Notice that all the levels are pushed up slightly with $E_1=0.1717$ (slightly more than 0).

Next, we explore the barrier-top ($E=V_0$) eigenvalues. For this we find the roots of the trigonometric Eqn. (25) to get the allowed values 
of $V_0$ for the fixed symmetric ($d_1=d_2=2,b=1)$ geometry of the potential (5).  We get  first four roots as ....$V_0=0.6168, 1.3098, 5.5516, 6.4693,...$ So the double-well with these heights of the in-barrier will have barrier-top eigenstates at these energies ($E=V_0$) with number of nodes as $n=0,1,2,3$, respectively. See Fig. 4, and the Table I. The linear part $(Bx+C)$ of the even eigenstates for $|x|<b$ is horizontal and slant for the odd ones.
We have checked that these barrier-top states (say, $\psi_b(x)$ of one potential is orthogonal
\begin{equation}
\int_{-a}^{c} \psi_B(x) \psi_n(x) dx =0.
\end{equation}
to all the 5 other  (listed in Table I) eigenstates of the same potential.

Again in order to check the robustness of these states we increase the barrier height $V_0$ from the special value 0.6168 to 0.7 to find that all the levels including $n=0$ are pushed up slightly. When we decrease the height $V_0$ from the special value 5.5516 to 5 one can see that all the levels including the  $n=2$ have been pushed down slightly. In Fig. 4, the even
barrier-top eigenstates are linear and horizontal in the barrier region ($x \in [-1,1]$), whereas the odd ones are linear and slant in $x \in [-1,1]$

Similarly, for the asymmetric case $(d_1=1,d_2=2,b=1)$ see the Table II and Fig. 3. Notice that the linear part  of the eigenstates $\psi_B(x)$ is only slant in the barrier region $x \in [-1,1]$ for all the four special rectangular double well potential.

The Figs. (6,7) aptly display the ordinary eigenstates for a special hole and a special double-well potentials emerging from 
the usual analysis as given in {\bf Case (ii)} above. Their special eigenstates which emerge from the {\bf Cases (i), (iii)} and make the spectra complete  are  displayed  in Figs. (2,4), respectively.
If these states are not included, the formidable oscillation theory [14] of Strum-Liouville eigenvalue problem will not be fulfilled. According to the oscillation theory the $n^{th}$ eigenstate has $(n-1)$ zeros (nodes). 

Lastly, we conclude that for a fixed geometry the usual rectangular double-well and the hole potentials become special for some calculable discrete values of the height and depth parameter. Then the double-well  entails the barrier-top eigenstate and the hole admits the zero-energy eigenstate. These eigenstates can be  detected only by invoking  the linear $(Bx+C)$ solution of Schr{\"o}dinger equation in the barrier region in the former case and out-side the hole in the latter case. But for these special eigenstates the spectrum of these 
special potentials would not be complete.

\renewcommand{\theequation}{A-\arabic{equation}}
\setcounter{equation}{0}
\section*{Appendix}

The Eqs.(9,10; 16,17; 22,23) are homogeneous equations in four unknowns: $(A,B,C,D)$. The may be written as
\begin{subequations}
\label{allequations}
\begin{eqnarray}
x_1 B+ y_1 C+ z_1 D= w_1 A \\ \label{equationa}
x_2 B+ y_2 C+ z_2 D= w_2 A \\ \label{equationb}
x_3 B+ y_3 C+ z_3 D= w_3 A \\ \label{equationc}
x_4 B+ y_4 C+ z_4 D= w_4 A 
\end{eqnarray}
\end{subequations}	
These  have a trivial solution
(0,0,0,0). However if the determinant,
\begin{equation}
\Delta=\left |\begin{array} {cccc}  x_1 & y_1 & z_1& w_1 
\\ x_2 & y_2 & z_2 & w_2 \\ x_3 & y_3 & z_3 & w_3 \\ x_4 & y_4 & z_4 & w_4 \\
\end{array} \right |=0,
\end{equation}
the Eqs. (A-1) can also have infinitely many non-zero (non-trivial) solutions. See the determinants in Eqs. (11,18,24). The first three of these equations can be solved by Cramer's method as
\begin{equation}
{B \over \Delta_1}= {C \over \Delta_2} = {D \over \Delta_3}={A \over \Delta_4},  
\end{equation}
where
\begin{equation}
\Delta_1=\left |\begin{array} {ccc} w_1 & y_1 & z_1 
\\ w_2 & y_2 & z_2\\ w_3 & y_3 & z_3 \\ 
\end{array} \right |, 
\Delta_2=\left |\begin{array} {ccc} x_1 & w_1 & z_1 
\\ x_2 & w_2 & z_2\\ x_3 & w_3 & z_3 \\
\end{array} \right |, 
\Delta_3=\left |\begin{array} {ccc} x_1 & y_1 & w_1 
\\ x_3 & y_2 & w_2\\ x_3 & y_3 & w_3 \\ 
\end{array} \right |, 
\Delta_4=\left |\begin{array} {ccc} x_1 & y_1 & z_1 
\\x_2 & y_2 & z_2\\ x_3 & y_3 & z_3 \\
\end{array} \right |, 
\end{equation}
When these values of $B,C,D$ are put in (A-1d) one recovers the consistency condition (A-2). Here we have taken $A$ as 1, however, one may take $A= N_n$ where 
$N_n$ is the normalization
constant for a given eigenstate  such that $N_n^2\int_{-\infty}^{\infty} \psi_n^2(x)=1$

\begin{table}

	\centering
	
		\begin{tabular}{|c||c||c||c||c||c||c||c|}
		\hline
		S.N. &$~~~~V_0~~~~$ & $E_0$ &  $E_1$ & $E_2$ & $E_3$ & $E_4$ & $E_5$ \\
		\hline
		\hline
		1 &0 & 0.2741& 1.0966 & 2.4674 & 4.3864& 6.8538&  9.8696 \\
		\hline
		2&10 & 1.8201& 1.8260 & 6.9444& 7.0626 & 
		11.7571&
		14.2955\\
				\hline
 		
		3&-5&-3.8520& -.8965& 1.5636& 2.8132& 5.3212& 8.5313\\
		\hline
		4&-0.4267& 0 & 1.0077 & 2.3368 & 4.2153 & 6.7348 & 9.7303 \\
		\hline
	   5&-3.3730& -2.3768 & 0 & 1.7942 & 3.1867 & 5.8598 & 8.9105\\
	   \hline
	  6& -10.8393& -9.4034 & -5.3272 & 0 & 2.1613 & 3.4687 & 7.2319 \\
	   \hline
	  7& -23.1923& -21.5077 & -16.5516 &-8.7346 & 0 
	  &2.3052 & 3.5647\\
	   \hline
	   8&-0.5&-0.0497& .9913 & 2.3165 & 4.1861 & 6.7142 & 9.7070\\
	   \hline
	  9& -3& -2.0483 & 0.1717 & 1.8469 & 3.2933 & 5.9776 & 9.0031\\
	   \hline

	           10&0.6168& $V_0$ &1.2069 & 2.7001 & 4.6341 & 7.0253 & 10.0817 \\
		\hline
		11&1.3098& 0.9193 & $V_0$ & 3.0287 & 4.9076 &7.2197 & 10.3359\\
		\hline
        12&5.5516 & 1.6280 & 1.6639 & $V_0$ & 6.2605 & 8.7725&
        12.2097\\  
        \hline
        13&6.4693 & 1.6836 & 1.7072 & 5.9698 & $V_0$ & 9.2801& 12.6555 \\ 
        \hline 
        14& 0.7 & 0.6577 & 1.2204 & 2.7358 & 4.6674 & 7.04851 & 10.1114 \\
        \hline 
        15& 5.0 & 1.5872 & 1.6342 & 5.2618 & 6.1215 & 8.5044 &
        11.9442\\
        \hline
		\end{tabular}
		\caption{First six eigenvalues of the symmetric (double well or hole) potentials (Fig. 1) when $V_0$ is varied. Here $b=1$ and $d_1=2=d_2$ ($a=3,c=3$).}
		\end{table}
		\begin{table}

	\centering
		\begin{tabular}{|c||c||c||c||c||c||c||c|}
		\hline
		S.N& $~~~~V_0~~~~$ & $n=0$ &  $n=1$ & $n=2$ & $n=3$ & $ n=4$ & $n=5$  \\
		\hline
		\hline
		1&0 & 0.3947 & 1.5791 & 3.5530 & 6.3165 & 9.8696 & 14.2122\\
		\hline
		2&10 & 1.8230 & 5.3753 & 7.0049 & 11.8832 & 14.9494 & 18.6187 \\
		\hline
				3&-5 & -3.840 & -0.7726 & 2.0291 & 4.5628 & 8.2361 & 12.1006 \\
		\hline
		4&-0.5695 & 0 & 1.3772 & 3.3188 & 6.1236 & 9.6476 & 13.9614 \\
		\hline
	   5&-3.7466 & -2.6896 & 0 & 2.3015 & 5.0178 & 8.59 27& 12.6006 \\
	   \hline
	   6&-11.2902 & -9.8388 & -5.6968 & 0 & 2.7129 & 6.5210 & 10.1178 \\ 
	   \hline
	   7&-23.6678& -21.9771 & -17.0001 & -9.1297 & 0 & 2.8689 & 7.6373\\
			   \hline
	  8& -0.5 & 0.0508 & 1.4012 & 3.3468 & 6.1472 & 9.6741 & 13.9920 \\
	    \hline
	  9& -3 & -2.0190 & 0.3995 & 2.4901 & 5.2842 & 8.8175 & 12.9108\\
		
		\hline
		10&0.8753& $V_0$ & 1.9470  & 3.9276 & 6.6146 &
		10.2338 & 14.5981
		 \\
		\hline
		11&3.2699& 1.4520& $V_0$ & 4.9066 & 7.5021 &
		11.3686 & 15.6440 \\
		\hline
       12& 6.1213& 1.6776 & 4.4754& $V_0$ &8.9904 & 12.8985& 16.8682\\  		
		\hline
		13 & 15.8550 & 1.9382 & 6.1100 & 7.6114 &  $V_0$ & 18.0552  & 21.7509\\
		\hline
14 & 1&.9289 & 2.0077& 3.9819& 6.6575 & 10.2880 & 14.6530\\
        \hline
15 & 3 & 1.4171 & 3.1233&4.8670 & 7.3917&11.2316 & 15.5272\\  
\hline      
		\end{tabular}
		\caption{First six eigenvalues of the asymmetric (double well or hole) potential (Fig. 1) when $V_0$ is varied. Here $b=1$ and $d_1=1,d_2=2$ $(a=2,c=3)$ }
		\end{table}
\begin{figure}		
\centering
\includegraphics[width=10 cm,height=5 cm]{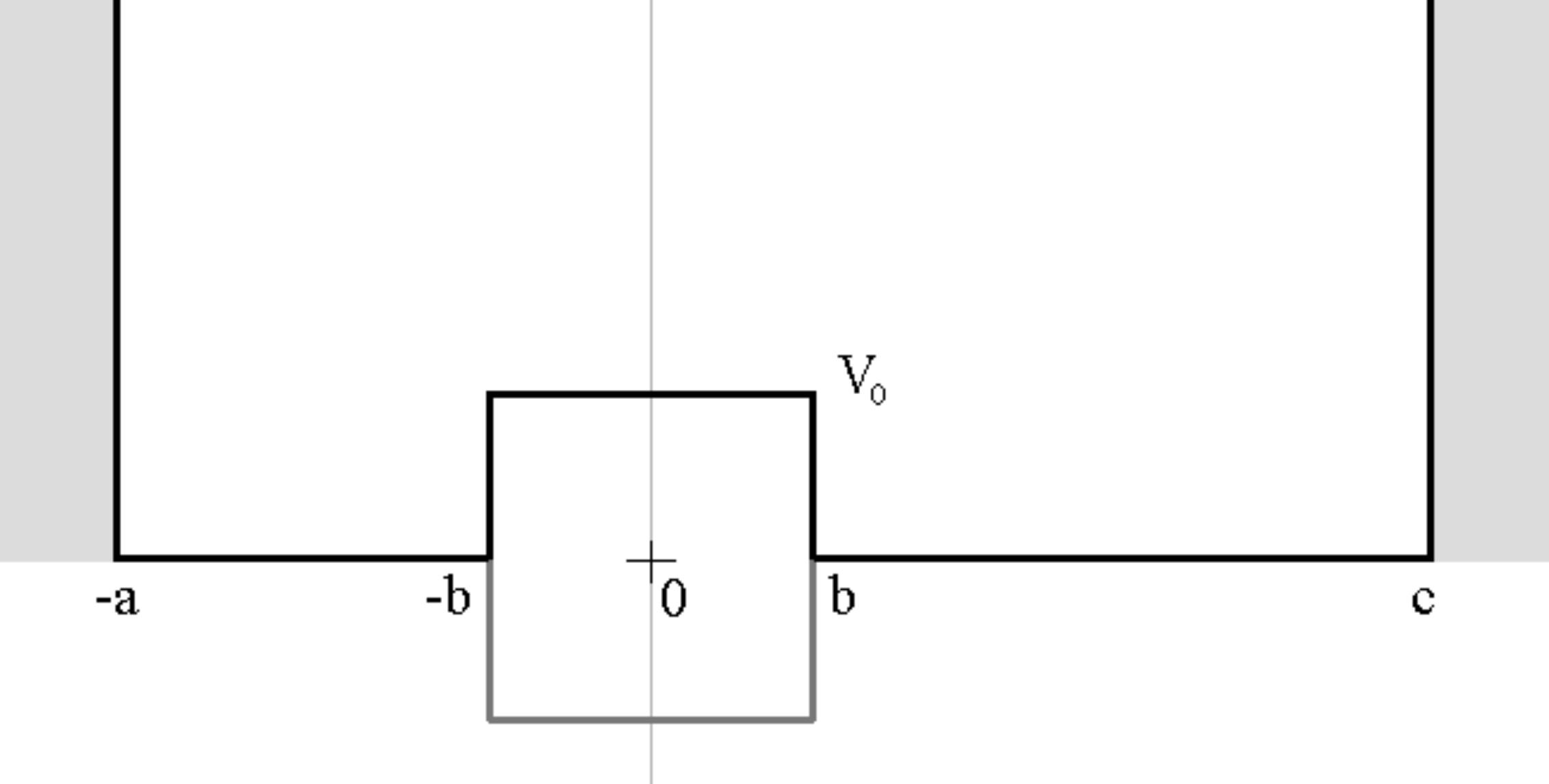}
\caption{Schematic depiction of the rectangular double-well (black line) and the rectangular hole (well/pit with gray line) potentials. The vertical long thin gray line denotes the Dirac Delta barrier/well at $x=0$. }	
\end{figure}	
\begin{figure}
\centering
\includegraphics[width=6 cm,height=3.5 cm]{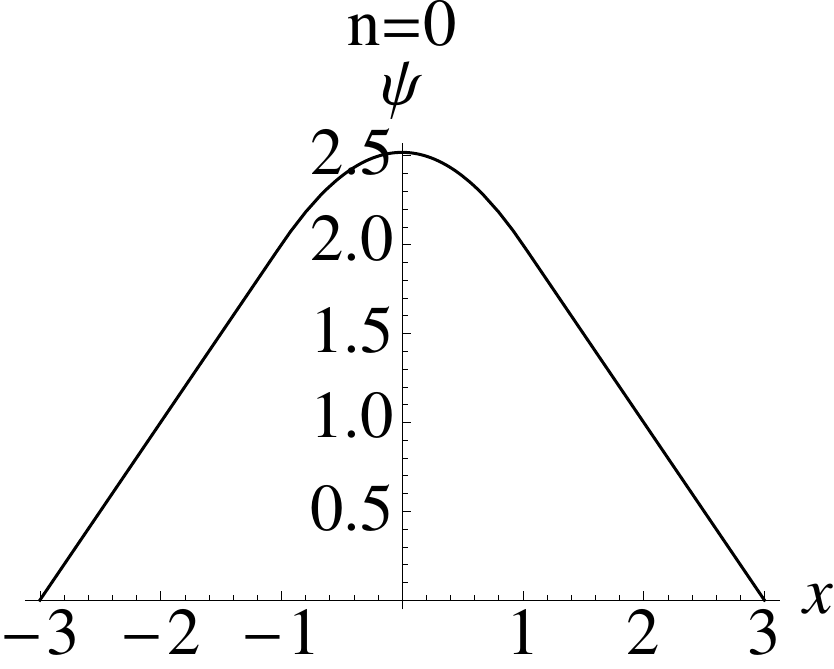}
\hskip .5 cm
\includegraphics[width=6 cm,height=3.5 cm]{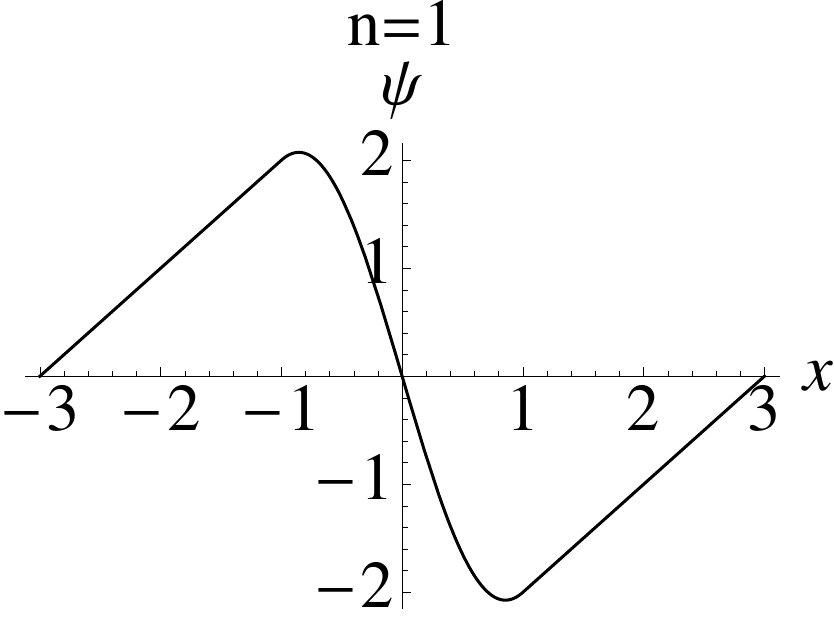}
\hskip .5 cm \\
\includegraphics[width=6 cm,height=3.5 cm]{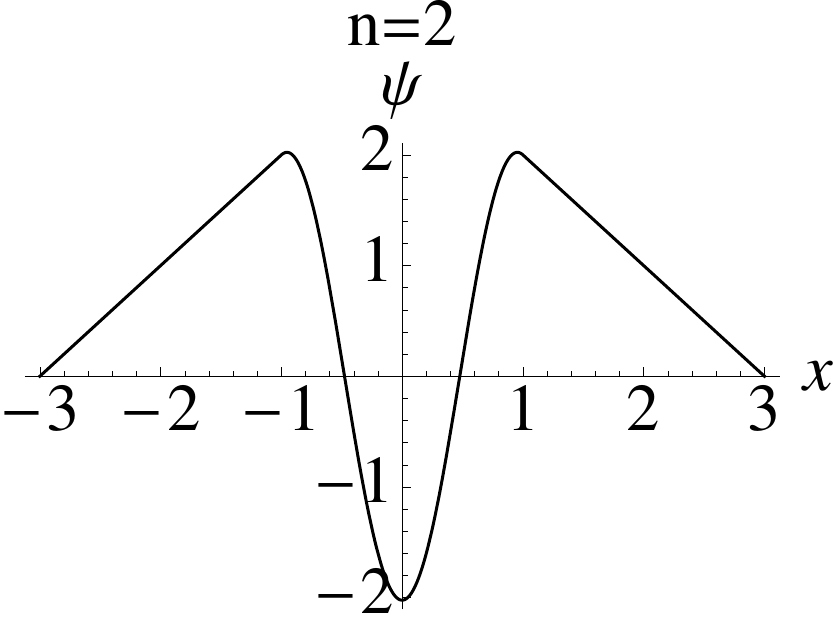}
\hskip .5 cm
\includegraphics[width=6 cm,height=3.5 cm]{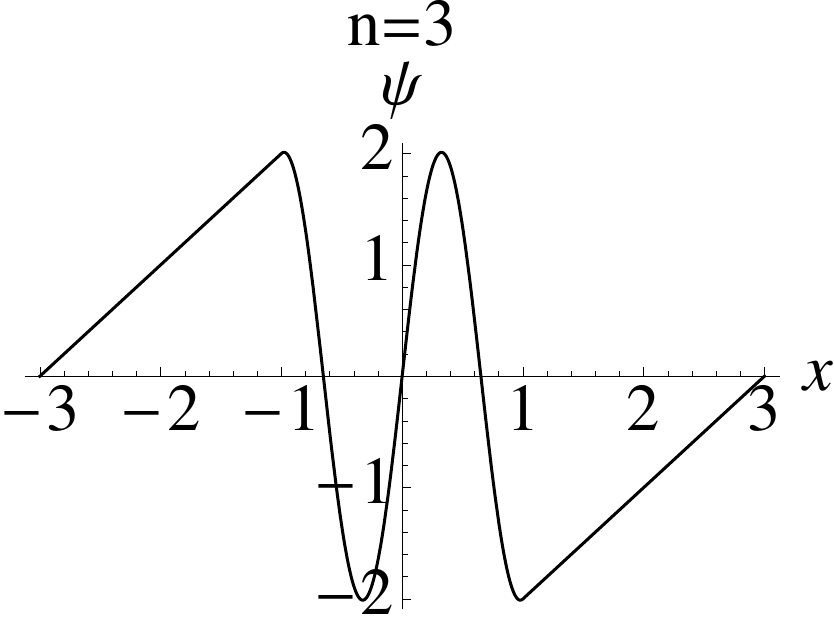}
\caption{The zero-energy un-normalized eigenstates, $\Psi_Z(x)$, of four symmetric hole potentials. See entries 4-7 of the Table I. Notice the linear ($Bx+C$) part for $|x|\ge 1$. All these functions are continuous and first-differentiable in the entire domain of $x \in [-3,3]$.}
\end{figure}
\begin{figure}
\centering
\includegraphics[width=6 cm,height=3.5 cm]{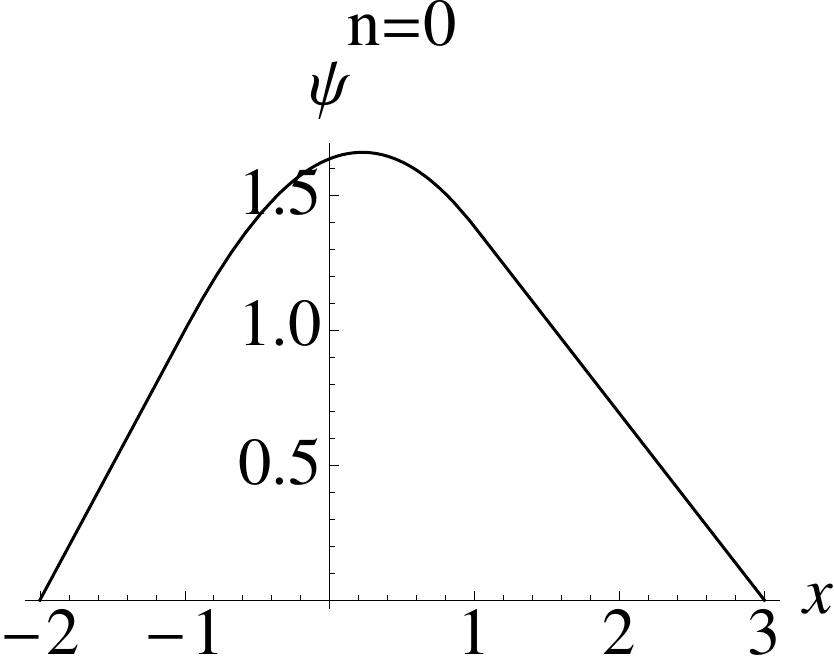}
\hskip .5 cm
\includegraphics[width=6. cm,height=3.5 cm]{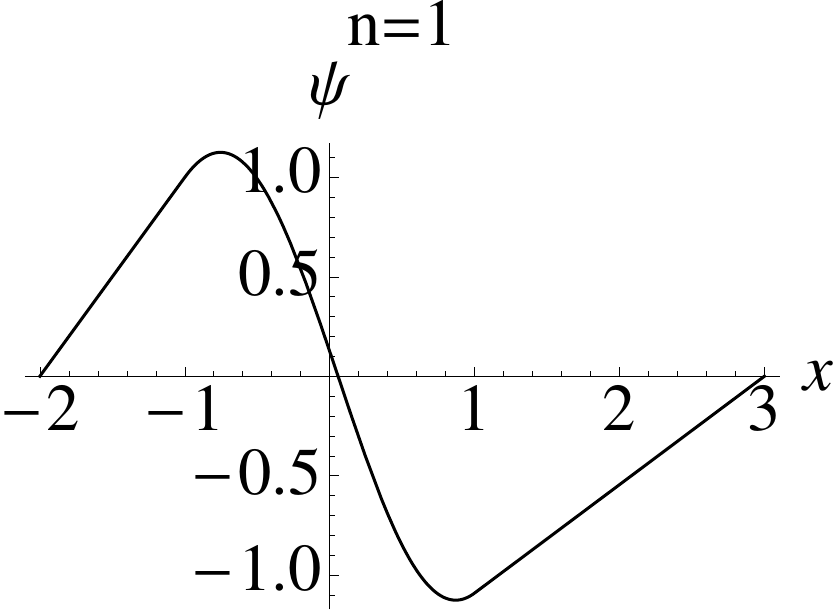}
\hskip .5 cm \\
\includegraphics[width=6.0 cm
,height=3.5 cm]{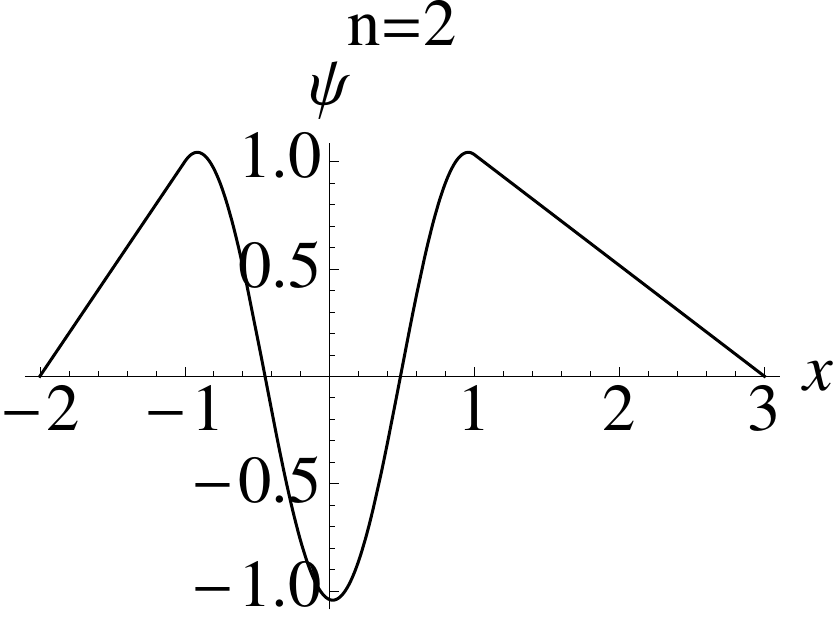}
\hskip .5 cm
\includegraphics[width=6 cm,height=3.5 cm]{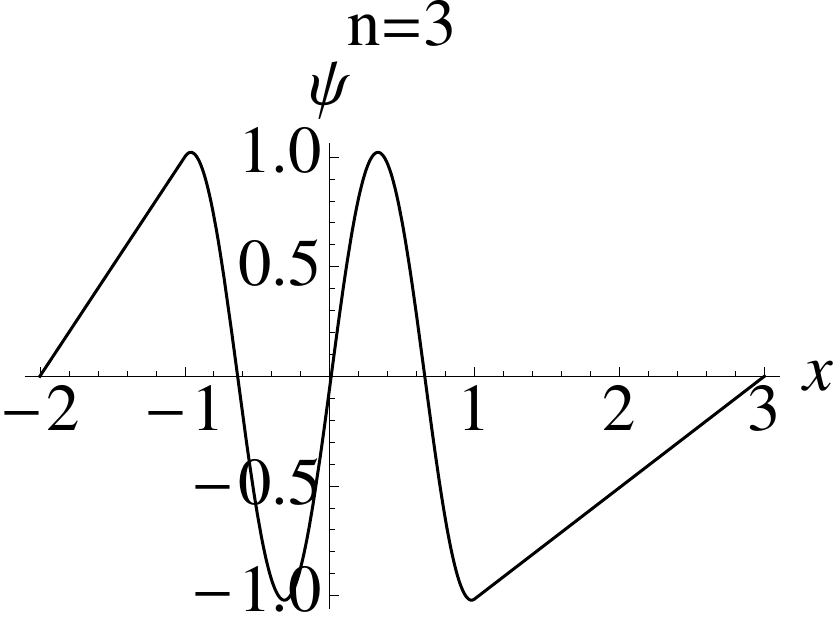}
\caption{The same as in Fig. 2, for the asymmetric
hole potential. See the entries 4-7 of the  Table II. }
\end{figure}
\begin{figure}
\centering
\includegraphics[width=6 cm,height=3.5 cm]{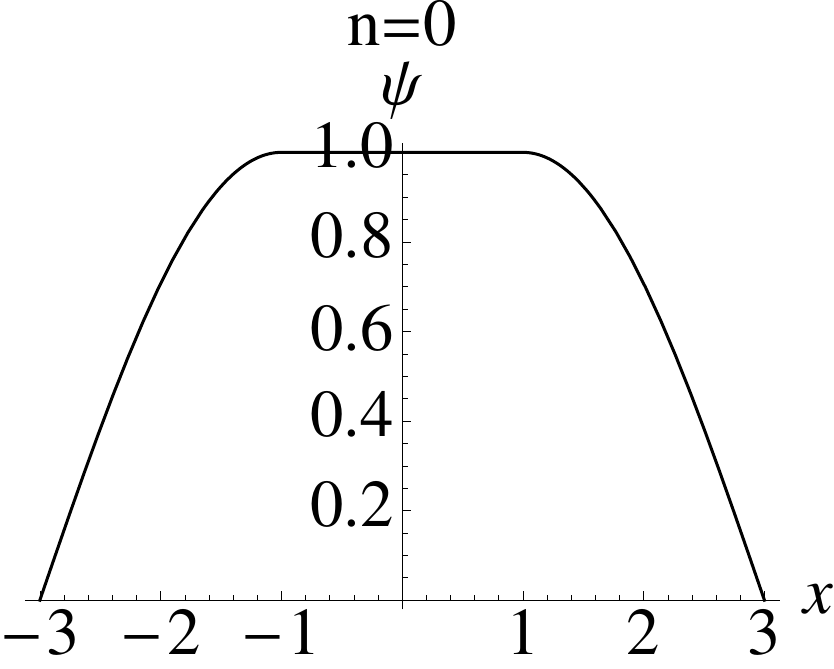}
\hskip .5 cm
\includegraphics[width=6 cm,height=3.5 cm]{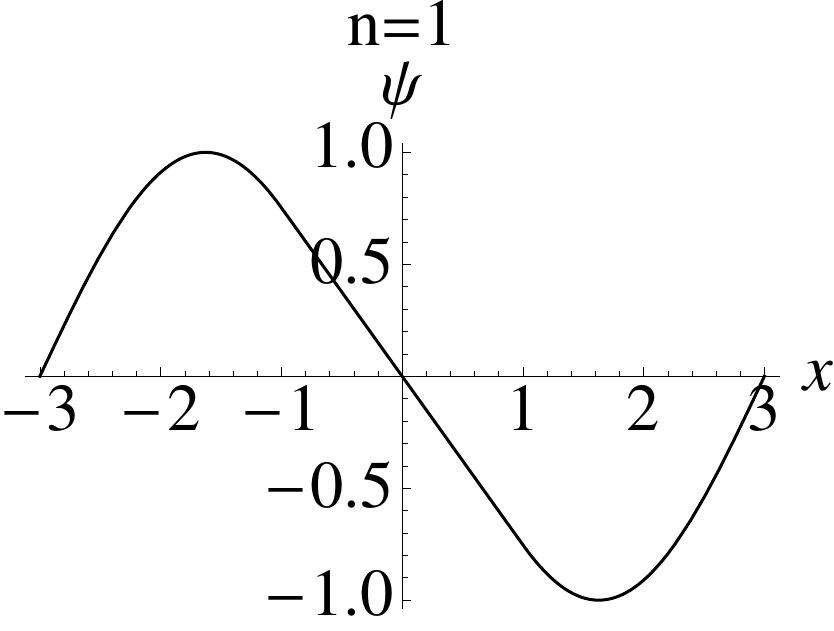}
\hskip .5 cm\\
\includegraphics[width=6 cm,height=3.5 cm]{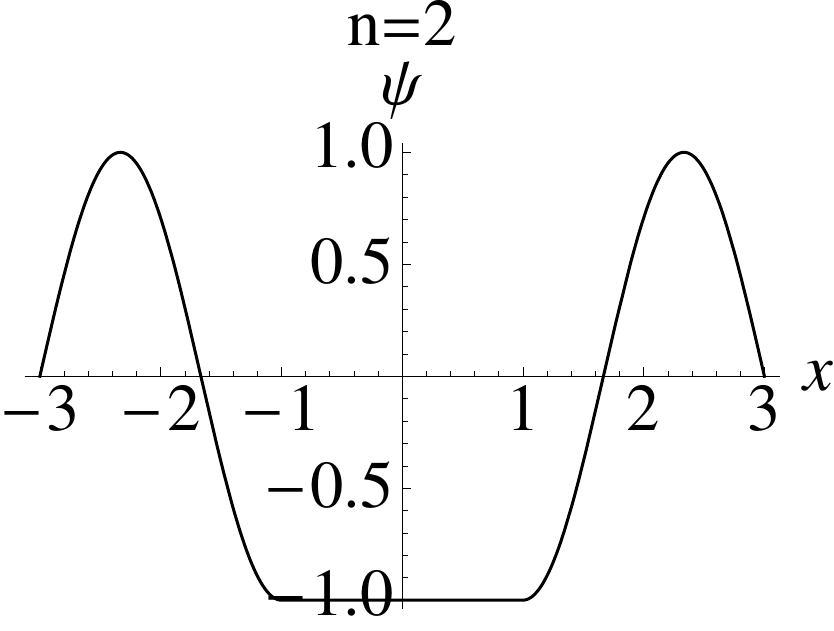}
\hskip .5 cm
\includegraphics[width=6 cm,height=3.5 cm]{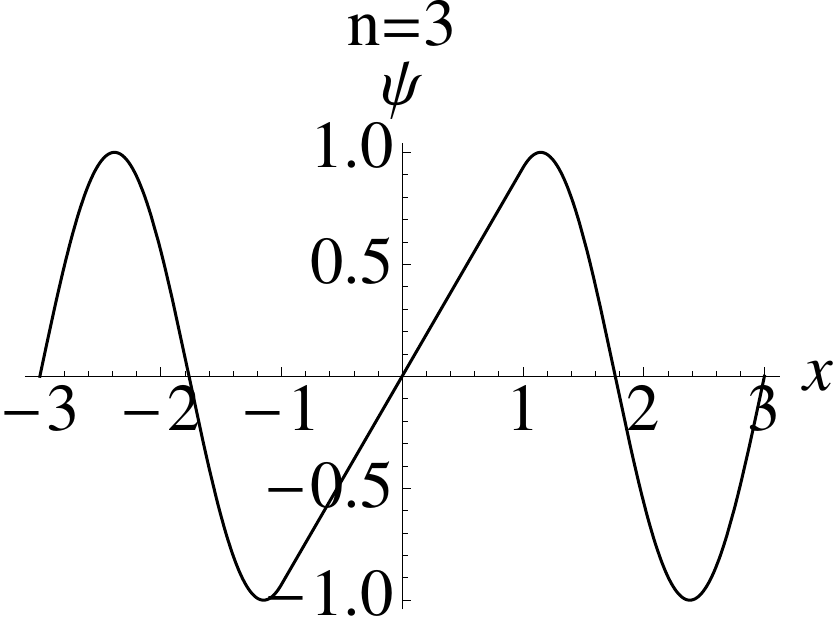}
\caption{The barrier-top un-normalized eigenstates, $\psi_B(x)$ of four potentials. Notice that this time the linear part $(Bx+C)$ is inside $x \in [-1,1]$. For the even state the linear parts are horizontal and slant for the the odd states. See the entries 10-13 of the Table I.}
\end{figure}
\begin{figure}
\centering
\includegraphics[width=6 cm,height=3.5 cm]{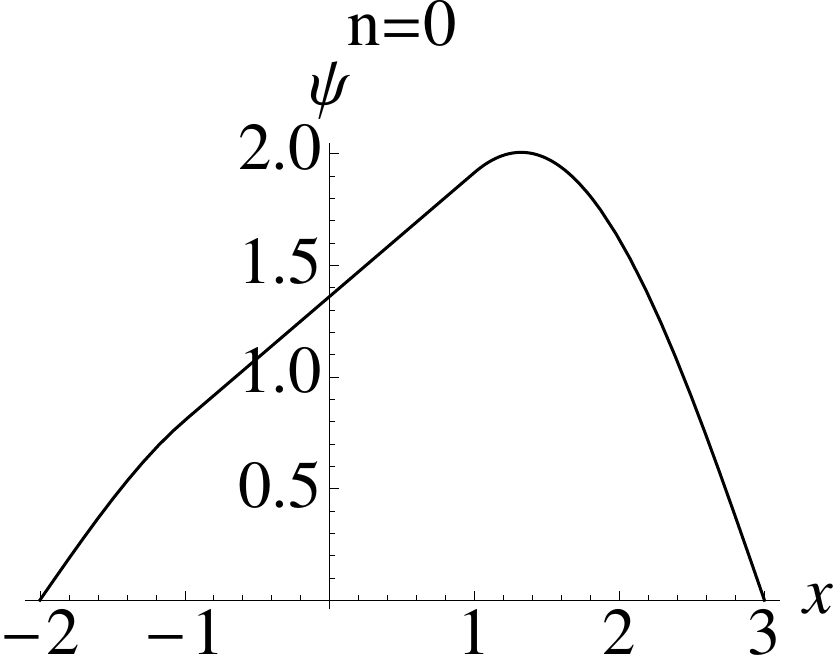}
\hskip .5 cm
\includegraphics[width=6 cm,height=3.5 cm]{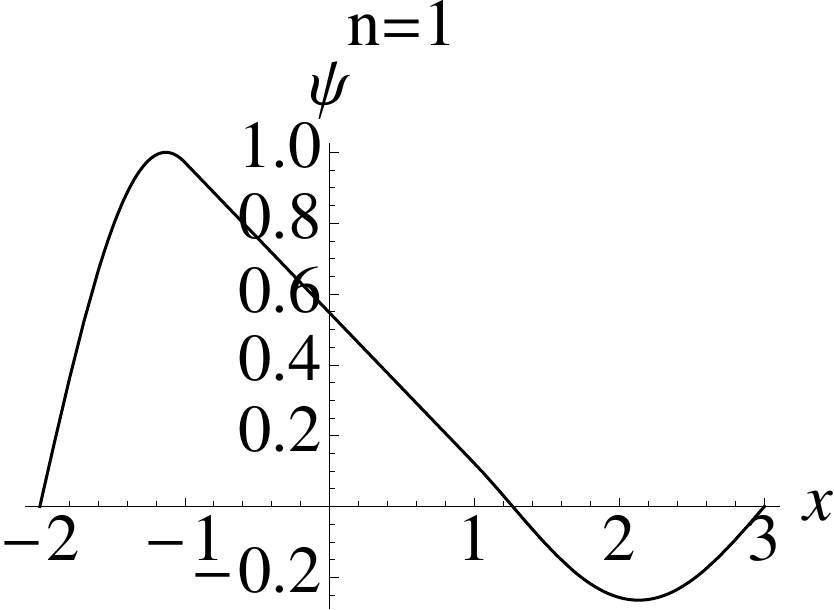}
\hskip .5 cm\\
\includegraphics[width=6 cm,height=3.5 cm]{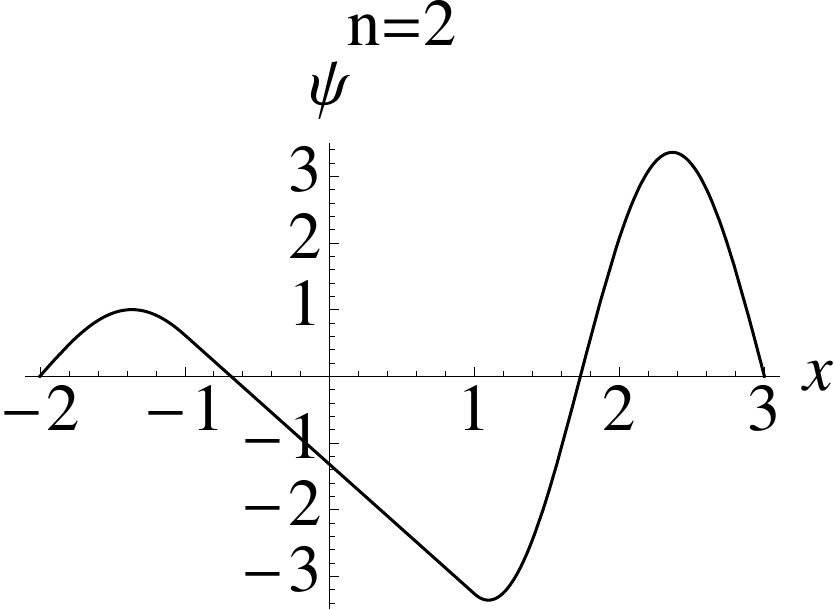}
\hskip .5 cm
\includegraphics[width=6 cm,height=3.5 cm]{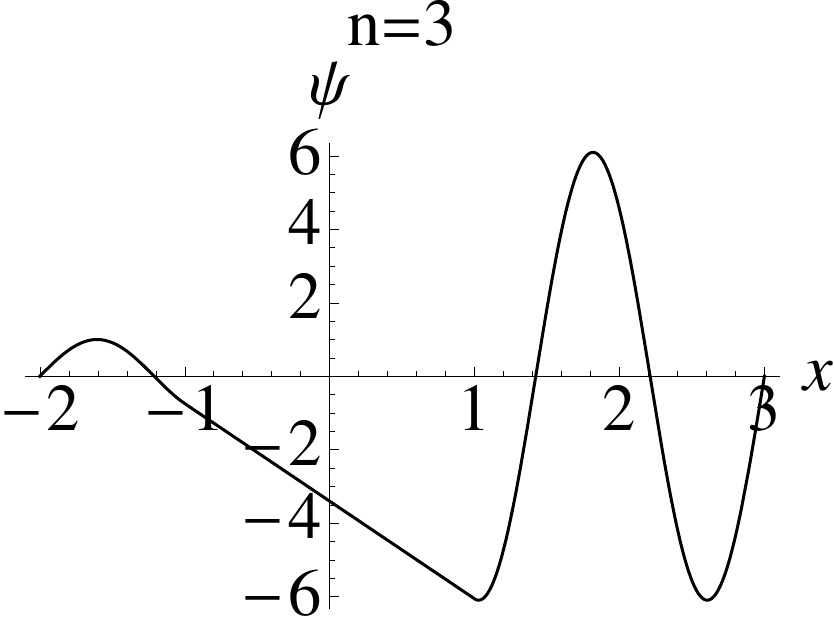}
\caption{The same as Fig. 4, for four asymmetric double-well potentials. This time the linear parts in all the cases is slant. See the entries 10-13 of the Table II.}
\end{figure}
\begin{figure}
\centering
\includegraphics[width=6 cm,height=3.5 cm]{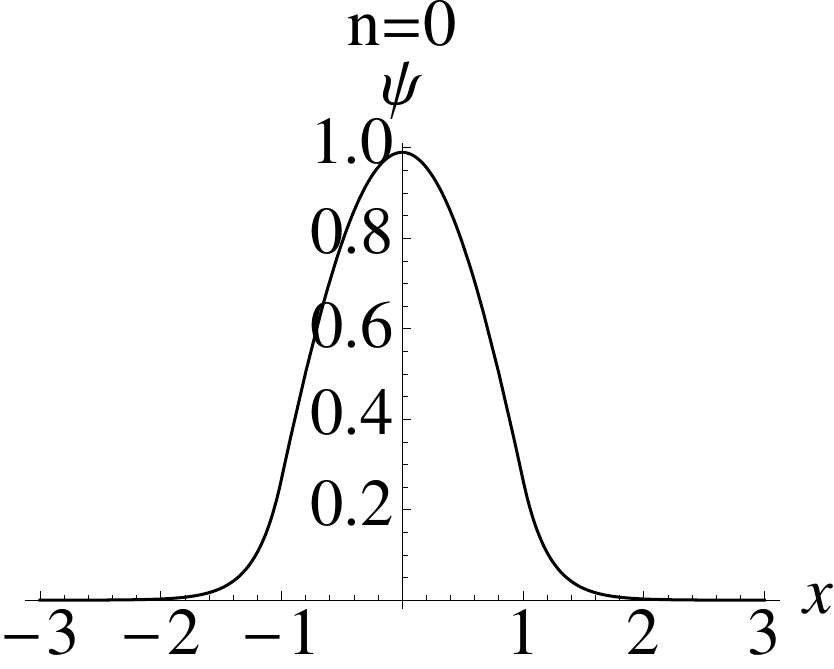}
\hskip .5 cm
\includegraphics[width=6 cm,height=3.5 cm]{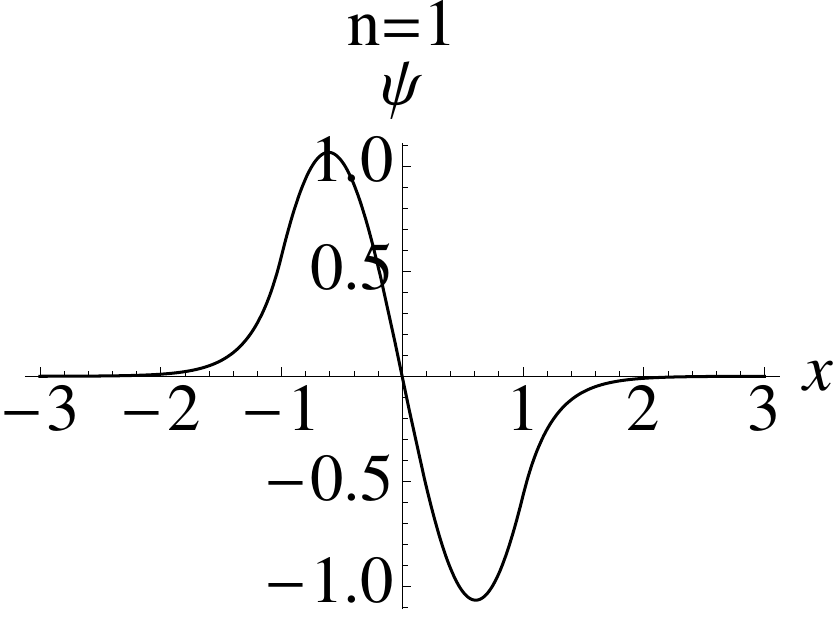}
\hskip .5 cm\\
\includegraphics[width=6 cm,height=3.5 cm]{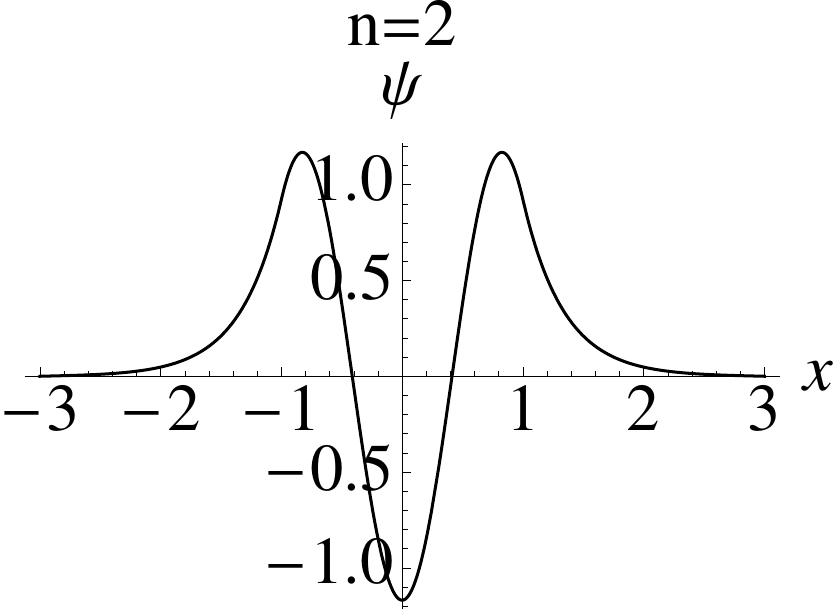}
\hskip .5 cm
\includegraphics[width=6 cm,height=3.5 cm]{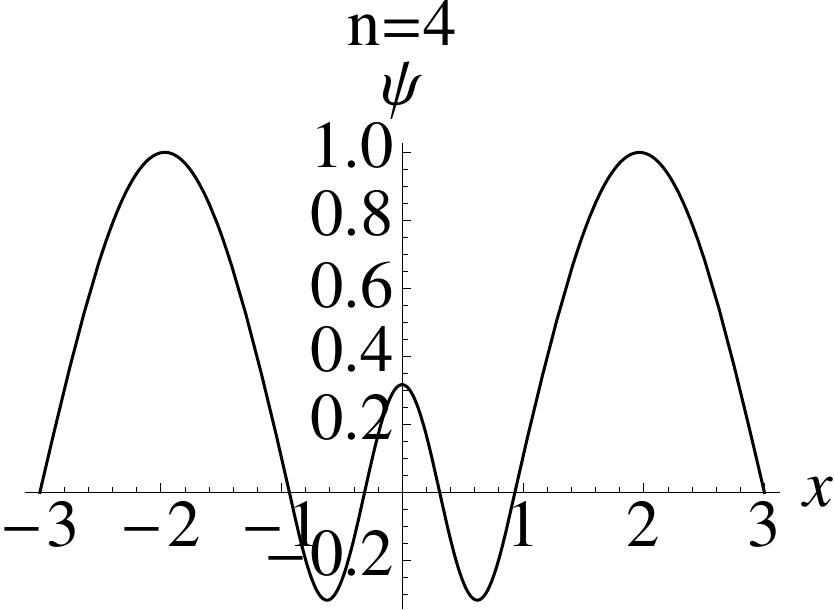}
\caption{The first four un-normalised eigenstates of the hole potential for the special value of $V_0=-23.1923.$ The special  $n=3$ (the zero energy) eigenstate of this symmetric potential (see the Table I) is given in Fig. 2. Notice that $n=4$ state has positive eigenvalue (2.0305) hence it does not fall of exponentially around the endpoints like those of $n=0,1,2$ eigenstates do.
}
\end{figure}

\begin{figure}
\centering
\includegraphics[width=6 cm,height=3.5 cm]{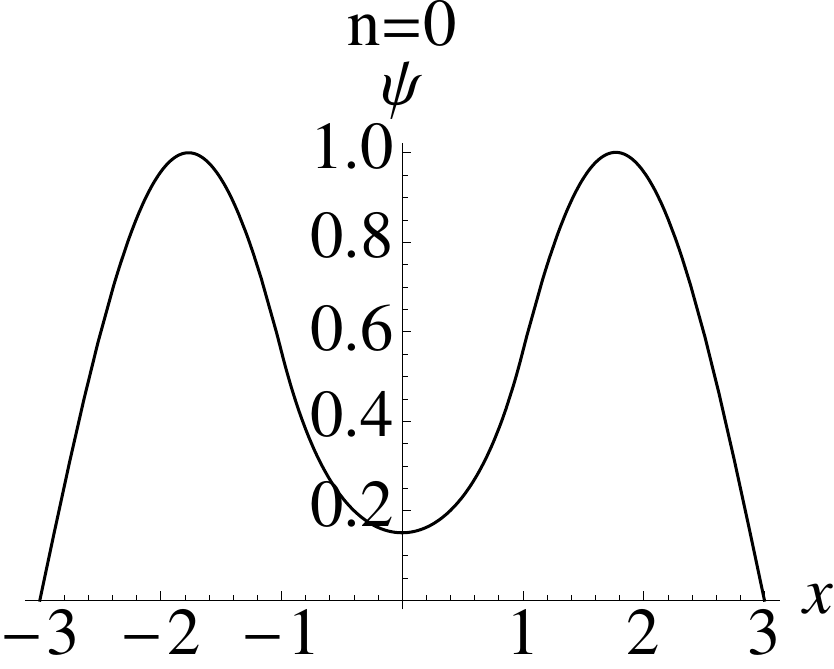}
\hskip .5 cm
\includegraphics[width=6 cm,height=3.5 cm]{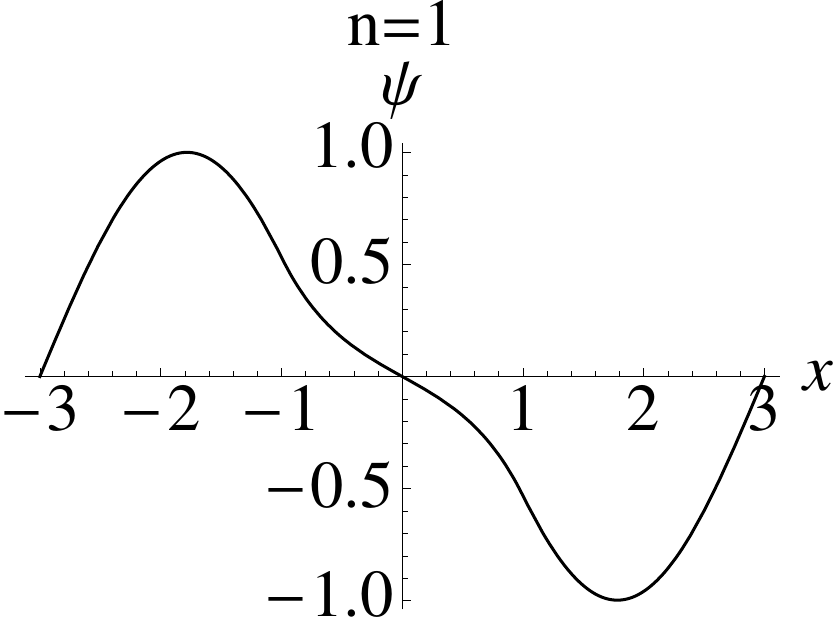}
\hskip .5 cm\\
\includegraphics[width=6 cm,height=3.5 cm]{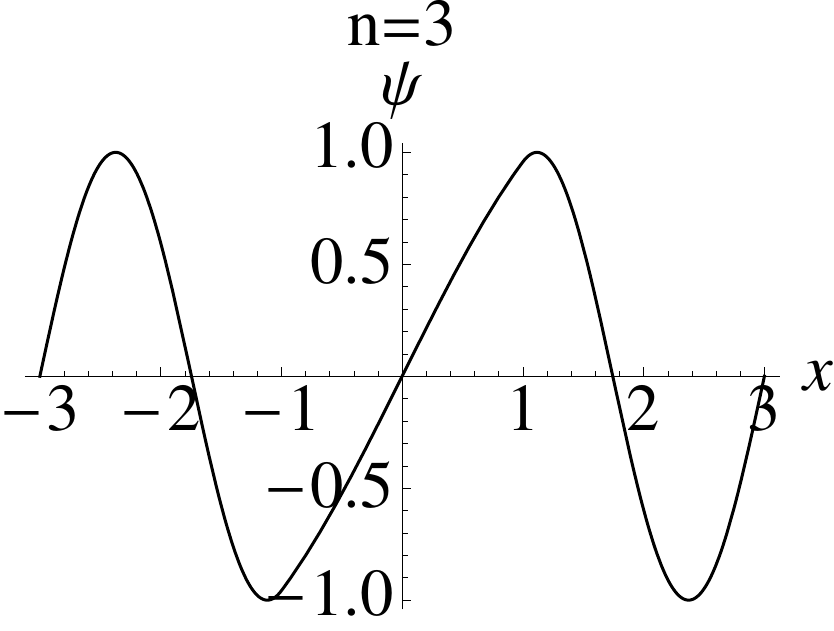}
\hskip .5 cm
\includegraphics[width=6 cm,height=3.5 cm]{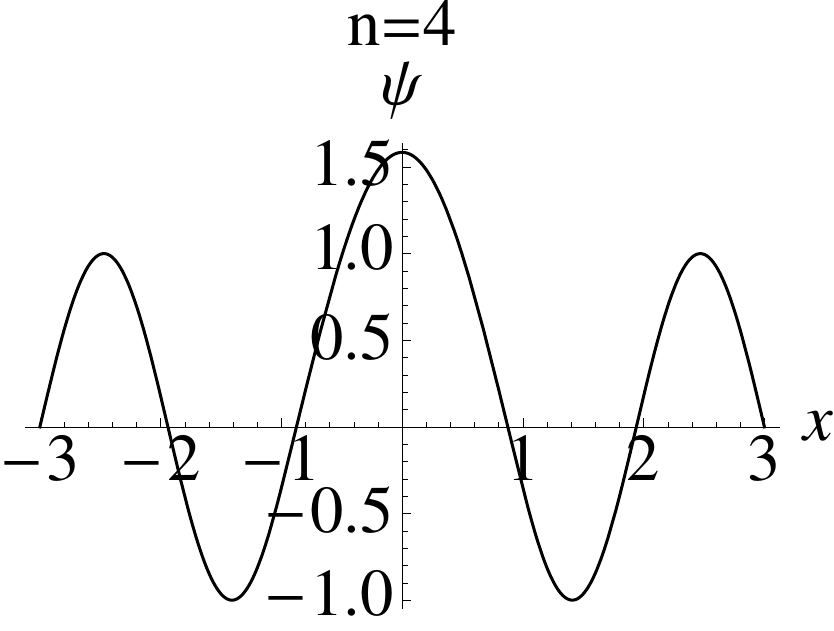}
\caption{The first four un-normalised eigenstates of the double-well potential for the special value of $V_0=5.5516$ The special  $n=2$ (the zero energy) eigenstate of this symmetric potential (see the Table I) is given in Fig. 4. Notice that the almost linear behaviour of $n=3$ eigenstate in $x \in [-1,1]$ is purely incidental.}
\end{figure}

\end{document}